# AN APPROACH OF SUBSTITUTION METHOD BASED ON ASCII CODES IN ENCRYPTION TECHNIQUE

*Avinash Sharma, Anurag Bhatnagar, Nikhar Tak, Anuradha Sharma, Jitendra Avasthi, Prerna sharma*

*Abstract -* In polyalphabetic substitution the plain text's letters are enciphered differently according to their position. The name polyalphabetic suggests that there are more than one key so we have used two keys combination instead of just one, in order to produce the cipher text. We can also use three or more keys to make the enciphering process more complicated. In this paper have produced ASCII Codes of the plain text and then we have reversed it said reverse ASCII Codes and then we have generated two keys named K1 and K2. K1 is generated by addition of reverse ASCII Codes and K2 is generated by addition of ASCII Codes. Then these K1 and K2 Keys are alternatively applied on Reverse ASCII codes in order to produce cipher text. On the Destination hand Deciphering is used to produce the plain text again. Our technique generates random cipher text for the same plain text and this is the major advantage of our technique.

## I. INTRODUCTION -

Sensitive information can't be sent over the internet without using security mechanism as this information may be accessed by unauthorized person in order to harm the message. So the demand for effective network security is increasing exponentially day by day. Encryption technique is used to transform the information over the internet. Encryption algorithm makes plain text into unreadable form, this unreadable form is known as cipher text, And this Cipher text transmitted through network and at the destination side the reverse technique called Decryption algorithm is used to convert from cipher text to plain text again.

## 2. HISTORY OF CRYPTOGRAPHY

Cryptography is a method or technique by which a message can be altered so that it becomes meaningless to anyone else but the intended recipient. This is done primarily in two basic ways, one is to change the position of letters or words within a message known as "Transposition", and the other is by substituting letters or words by different ones, known as "Substitution" respectively. The word cryptography comes from the Greek word **kryptos,** which means **hidden** and **graphein**, which means **writing**. Cryptography, the science of encrypting and decrypting information can be traced back all the way to year 2000 BC in Egypt.

The history of cryptography can be broadly divided into three phases -

2.1 According the first recorded use of cryptography for correspondence was by the Spartans who (as early as 400 BC) employed a cipher device called a **"scytale"** to send secret communications between military commanders. The scytale consisted of a tapered baton around which was wrapped a piece of parchment inscribed with the message. Once unwrapped the parchment appeared to contain an incomprehensible set of letters, however when wrapped around another baton of identical size the original text appears.

2.2 During the second World Wars was inventing the **rotor cipher machine** for cryptography which provide both mechanical and electromechanical technology. These cipher text were hard to break but by the time they turned weaker and it becomes easy to break them.

2.3 After the era of those rotor ciphers and the World War 2 the electronics that had been developed in support of radar were adapted to crypto machines. The first electrical crypto machines were little more than rotor machines where the rotors had been replaced by electronic substitutions which use at this time for cryptography techniques.

### 3. SUBSTITUTION TECHNIQUES -

In this technique the letters in plain text are replaced by other letters, symbols or numbers. This makes plain text changed and non-understandable for others.

There are different types of substitution cipher:

3.1 *Monoalphabetic substitution cipher* - It uses fixed substitution over the entire message.
Example -

Plain Text : a b c d e a
Cipher Text : E F G H I E

Here is $5 \times 10^{26}$ possible keys to replace plain text. The problem with monoalphabetic substitution cipher is, it is easy to break because for the same plaintext it always produces the same letter of cipher text so we have limitation to replace plain text this makes cryptanalysis is easier.

3.2 *Polyalphabetic substitution cipher* – It uses a number of substitutions at different positions in the message. In this technique a set of related monoalphabetic substitution rules is used and a key is used that determines which particular rule is chosen for a given transformation. Merit of this technique is, different cipher texts are produced for the same plain text.
Example –

Key : a b c d a b c
Plain Text : w e l c o m e
Cipher Text : x g o g p o h

3.3 *Other Substituion techniques are* -

(a.) Simple substitution
(b.) Homophonic substitution
(c.) Polygraphic substitution

### 4. A MODEL OF SYMMETRIC AND ASYMMETRIC KEY CRYPTOGRAPHY -

4.1 Symmetric Encryption - This system uses only private keys. This requires the private key (code) to be installed on specific computers that will be used for exchanging messages between certain users. The system works pretty much like two best friends using a decoder ring to send secret messages to each other. Both friends know which code they are using and thus, only they will have the key to crack and encode secret messages. So same key is used both the side to encrypt and decrypt data, thus this is known private key encryption or symmetric encryption.

4.2 Asymmetric Encryption - The Asymmetric Encryption system uses both the private and public keys. The private key is for yourself and the public key is published on line for others to see. They use the public key to access the encrypted code that corresponds to your private key. Ex - If Hari is sending an encrypted message to Gopal which he does not want others to see, Hari would use his public key to encrypt it, and Gopal will be able to decrypt it with his own corresponding private key. Likewise, if Gopal sends a message to Hari, he will use Hari's public key to encrypt the message and Hari would use his own private key to decrypt it.

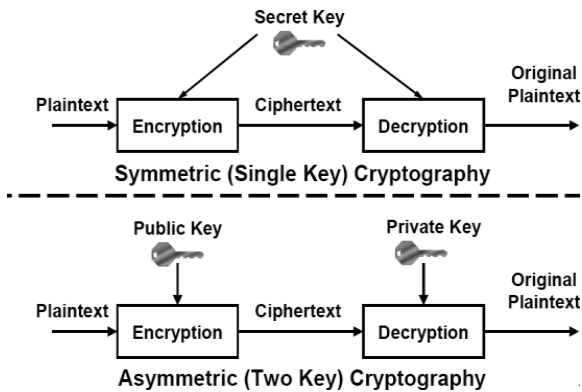

Figure - 3.1 Symmetric and Asymmetric Cryptography

### 5. ENCRYPTION PROCESS -

- For the encryption process as shown in Table – 7.1. First we produced ASCII Codes of the plain text and then we reversed it said reverse ASCII Codes.
- Take the example text "RESPECTEVERYONE"
- After that generated two key K1 and K2 and assign the values. K1 is generated by addition of reverse ASCII Codes and K2 generated by addition of ASCII Codes. For given example

$$K1=1056; K2 = 1155$$

Then these K1 and K2 Keys are alternatively applied on Reverse ASCII Codes in order to produce Cipher Text (Encrypted message). It generated a combination of 2*(256*256) letters encrypted coded text with 128 bit manner.

### 6. DECRYPTION PROCESS -

Decryption process is a reverse technique of encryption. So in this process subtract key value of all given text, the resultant text is a decrypted message and it generated a combination of 2*(256*256) letter decrypted coded text.

- In this process first subtract value of K1 from the first value of cipher text and value of K2 from the second value, alternatively subtract the value of K1, K2 to consecutive. Resultant text is a reverse ASCII code of each alphabet.
- After that reverse it to find decrypted message as "RESPECTEVERYONE" and this process shown in Table – 7.2.

Table – 7.1 Encryption Result

| Cipher Text | Reverse ASCII Code | ASCII Code | Plain Text |
|---|---|---|---|
| 1084 | 28 | 82 | R |
| 1251 | 96 | 69 | E |
| 1094 | 38 | 83 | S |
| 1163 | 08 | 80 | P |
| 1152 | 96 | 69 | E |
| 1231 | 76 | 67 | C |
| 1104 | 48 | 84 | T |
| 1251 | 96 | 69 | E |
| 1124 | 68 | 86 | V |
| 1251 | 96 | 69 | E |
| 1084 | 28 | 82 | R |
| 1253 | 98 | 89 | Y |
| 1153 | 97 | 79 | O |
| 1242 | 87 | 78 | N |
| 1152 | 96 | 69 | E |

## 7. RESULTS

Let's suppose that we need to transmit the plain text say REPECTEVERYONE, as shown below.

Plain Text = RESPECTEVERYONE

Now we will first find the ASCII codes for each of the letter in the plain text, as shown in the table 7.1. Then we will find two keys lets says K1 and K2.

K1 is the sum of reverse ASCII codes of the plain text.

K2 is the sum of ASCII codes of the plain text.

K1=1056; K2 = 1155

Now these keys are applied on the plain text in order to find out the cipher text.

K1 and K2 are alternatively added in the ASCII codes and cipher text is produced as show below.

Cipher Text = (1084,1251,1094,1163,1152,1231,1104,1251, 1124,1251,1084,1253,1153,1242,1152)

Two or more keys can be used in order to make enciphering and deciphering procedure more complex so it will become harder for unauthorized persons to analyze the original message. As shown in Table -1, For the plain text 'E' have the cipher text 1251 three times, and 1152 two times, so it behaves randomly in generating cipher text for the plain text and make it harder to break by grouping and guessing on the basis of the same cipher text values.

| Plain Text | ASCII Code | Reverse ASCII Code | Cipher Text |
|---|---|---|---|
| R | 82 | 28 | 1084 |
| E | 69 | 96 | 1251 |
| S | 83 | 38 | 1094 |
| P | 80 | 08 | 1163 |
| E | 69 | 96 | 1152 |
| C | 67 | 76 | 1231 |
| T | 84 | 48 | 1104 |
| E | 69 | 96 | 1251 |
| V | 86 | 68 | 1124 |
| E | 69 | 96 | 1251 |
| R | 82 | 28 | 1084 |
| Y | 89 | 98 | 1253 |
| O | 79 | 97 | 1153 |
| N | 78 | 87 | 1242 |
| E | 69 | 96 | 1152 |

Table – 7.2 Decryption Result

The Plain text is = RESPECTEVERYONE

8. KEY TERMS

**Block –** It's simply the input plain text which is to be encrypted. A sequence of consecutive characters that are encoded to transmit it to destination end.

**Block length –** It can be defined as the number of characters in a block is known as a block length.

**Key** - A relatively small amount of information that is used by an algorithm to customize the transformation of plaintext into cipher text (during encryption) or vice versa (during Decryption).

**Key length** - The size of the key - how many values comprise the key? The larger key length provides higher security and it makes cipher break hard to break.

**Chromosome** - The genetic material of an individual represents the information about a possible solution to the given problem.

**Plain text** - A message before encryption or after decryption, i.e., in its usual form which anyone can read, as opposed to its Encrypted form.

**Cipher text** - The result of encryption process is unreadable and non-understandable form, is known as cipher text.

**Encryption Algorithm** - An algorithm for performing encryption (and the reverse, decryption) - a series of well-defined steps that can be followed as a procedure. Works at the level of individual letters, or small groups of letters.

**Decryption Algorithm** - An algorithm for performing decryption (and the reverse, encryption) - a series of well-defined steps that can be followed as a procedure. Works at the level of individual letters, or small groups of letters.

**Cryptanalysis** - The analysis and deciphering of cryptographic writings or systems. Or it is also known as a procedure of breaking of cipher text into plain text.

**Cryptography** - The process or skill of communicating in or deciphering Secret writings or ciphers. *Cryptography* can be defined as the conversion of data into a scrambled code that can be deciphered and sent across a public or private network.

**Cryptosystem** - The package of all processes, formulae, and instructions for encoding and decoding messages using cryptography.

**Encryption/Enciphering** - The process of putting text into encoded form to make them unreadable and un-understandable to unauthorized user.

**Decryption/Deciphering** - Any procedure used in cryptography to convert cipher text (encrypted data) into plaintext.

**Mono alphabetic** - Using one alphabet - refers to a cryptosystem where each alphabetic character is mapped to a unique alphabetic character. The mono-alphabetic substitution cipher is so called because each plain text letter is substituted by the same cipher text letter throughout the entire message, for example if the key is 4, then plaintext 'a' will always be replaced by cipher text 'D'.

**Polyalphabetic** - Using many alphabets - refers to a cipher where each alphabetic character can be mapped to one of many possible alphabetic characters.

**Transposition -** In cryptography, a **transposition cipher** is a method of encryption by which the positions held by units of plaintext (which are commonly characters or groups of characters) are shifted according to a regular system, so that the cipher text constitutes a permutation of the plaintext. That is, the order of the units is changed. Mathematically a injective function is used on the characters' positions to encrypt and an inverse function to decrypt.

**Substitution -** In cryptography, a **substitution cipher** is a method of encryption by which units of plaintext are replaced with cipher text, according to a regular system, the "units" may be single letters (the most common), pairs of letters, triplets of letters, mixtures of the above, and so forth. The receiver deciphers the text by performing an inverse substitution. Substitution ciphers can be compared with transposition ciphers. In a transposition cipher, the units of the plaintext are rearranged in a different and usually quite complex order, but the units themselves are left unchanged. By contrast, in a substitution cipher, the units of the plaintext are retained in the same sequence in the cipher text, but the units themselves are altered.

## 9. CONCLUSION -

There are many approaches such as RSA algorithm, IDEA algorithm, AES algorithm, DES algorithm, DIFFIE-HELLMAN algorithm and many more that can be used to convert a plain text into cipher text to transmit over the network so nobody else than an intended recipient can understand the message. But Substitution and Transposition is the base for every algorithm as each and every algorithm uses Transposition or Substitution or both of these methods. In this regard we have introduced a new approach that is named as substitution using ASCII Codes. This new methodology for text encryption and decryption behaves randomly so grouping of the same cipher text and breaks it by just guessing it becomes more difficult as shown in Table – 7.1.

## 10. REFFERENCES -

Colorado State University, Fort Collins, CO 80523.

## 11. ABOUT THE AUTHORS

*Mr. Avinash Sharma is working as Associate Professor at engineering college affiliated to Pune Universiy. To his credit, he has more than 110 paper published in various conferences proceedings of national and international conferences and also in few international journals. He is guiding many research scholars for graduate and postgraduate degree programmes of reputed university. He has approx.15 year experience. Being a graduate from Mumbai University and post graduate from BITS, Pilani, the leading university of Asia, he had guided many research projects and award winning projects for ROBOTICs under Linux for You Magazine. He published four text books with local publishers and also working for government sponsored projects under Ministry of Educations for Schools and colleges. His research interest is in areas of e-learning, software engineering, advance database management systems etc.*

*Mr.Anurag Bhatnagar is a Research Scholar pursuing M.Tech from Rajasthan Technical University, KOTA.*

*Mr.Nikhar Tak is a Research Scholar pursuing M.Tech.*

*Ms. Anuradha Sharma is a Research Scholar pursuing M.Tech.*

*Mr,Jitendra Avasthi is a Research Scholar pursuing M.Tech from Rajasthan Technical University, KOTA.*

*Ms.Prerna Sharma is a Research Scholar pursuing M.Tech from Rajasthan Technical University, KOTA.*